\documentclass[conference]{IEEEtran}
\usepackage{amsmath,bm,amssymb}
\usepackage{stfloats}
\usepackage{graphicx}
\usepackage{multirow}
\usepackage{balance}
\def\BibTeX{{\rm B\kern-.05em{\sc i\kern-.025em b}\kern-.08em
    T\kern-.1667em\lower.7ex\hbox{E}\kern-.125emX}}

\begin{document}

\title{Source Optimization in MISO Relaying  \\with Channel Mean Feedback: \\A Stochastic Ordering Approach}
\author{Minhua Ding, {\em Member, IEEE},  and Q. T. Zhang, {\em Fellow, IEEE} \\Department of Electronic Engineering, City University of Hong Kong\\Email: minhua.ding@ieee.org, eekzhang@cityu.edu.hk}

\maketitle
\begin{abstract}
This paper investigates the optimum source transmission strategy to maximize the capacity of a multiple-input single-output (MISO) amplify-and-forward relay channel, assuming source-relay channel mean feedback at the source. The challenge here is that relaying introduces a nonconvex structure in the objective function, thereby excluding the possible use of previous methods dealing with mean feedback that generally rely on the concavity of the objective function. A novel method is employed, which divides the feasible set into two subsets and establishes the optimum from one of them by comparison. As such, the optimization is transformed into the comparison of two nonnegative random variables in the \textit{Laplace transform order}, which is one of the important stochastic orders.
It turns out that the optimum transmission strategy is to transmit along the known channel mean and its orthogonal
eigenchannels. The condition for rank-one precoding (beamforming) to achieve capacity is also determined.
Our results subsume those for traditional MISO precoding with mean feedback.
\end{abstract}

\section{Introduction\label{intro}}
Wireless relay systems have been a recent subject of intensive research due to their potential of  providing increased diversity, extended coverage, or flexibility  in compromising system performance and complexity/power consumption. Efficient relaying protocols have been developed, among which the most popular are the amplify-and-forward (AF) and decode-and-forward protocols~\cite{Laneman_1}.

To further enhance system performance, multiple antennas are deployed on one or more nodes of cooperative relay networks~\cite{Nabar_performance}-\cite{Dharmawansa_McKay_journal}. However, in multi-antenna systems, the optimum system structure and the resultant performance (capacity, error rate, mean-square error, etc.) depend heavily on the nature of channels and the channel state information (CSI) available at the transceiver (see~\cite{MIMO_capacity_goldsmith}-\cite{mding_sdb_T_IT_2010_to_appear} and references therein). Therefore, research on multi-antenna relay communications has progressed from early work assuming perfect CSI~\cite{tang_hua_perfect_CSI}\cite{medina_Olga} to recent studies assuming more realistic partial CSI~\cite{Prathapa_ICC}.

Consider a \textit{half-duplex} AF relay link with \(M\) antennas (\(M\geq 2\)) at the source and a single antenna at both the relay and the destination. This scenario typically occurs when a traditional multiple-input single-output (MISO) link (without relaying) is obstructed, and then a relay is used to maintain the link and coverage. We assume that the destination has full knowledge of the source-relay and relay-destination channels, whereas the relay and the source are only aware of the long-term statistics of the source-relay channel. Relaying without instantaneous CSI is commonly referred to as \textit{noncoherent relaying}.

The above scenario has been considered in~\cite{Prathapa_ICC}, where the source and the relay are assumed to have the long-term \textit{covariance information} of a rapidly changing source-relay channel, and the optimum source covariance to maximize the ergodic capacity is determined. However, when the \textit{mean information}\footnote{Channel covariance (mean) information at the transmitter is also referred to as channel covariance (mean) feedback~\cite{MIMO_capacity_goldsmith}.} of a slowly varying source-relay channel is available at the source, the capacity-achieving transmission strategy remains as an open problem and is tackled in this work. Mean feedback  models the uncertainty in CSI due to channel estimation errors, quantization errors or imperfectness of the feedback link, and is feasible in slowly varying fading channels~\cite{Zhou_Giannakis_CMF}.

Channel mean and covariance feedback generally require different treatments, as they constrain the solution differently. Regarding utilizing mean feedback for maximum ergodic capacity in traditional MISO, the optimization method in~\cite{Visotsky_madhow}~\cite{Jafar_Goldsmith} is based on calculus of variations using  the Fr\'{e}chet differential, whereas in~\cite{ALMoustakas}, the expression for ergodic capacity is first obtained and then ordinary calculus is used for optimization. Both methods rely on the concavity of the objective function in transmit covariance~\cite{Visotsky_madhow}-\cite{ALMoustakas}. However, such concavity cannot be established in our case, and these methods do not apply.

We, therefore, take a nonconventional powerful approach based on one of the stochastic orders,
the \textit{Laplace transform order}~\cite{Shaked_Shanth}, which circumvents the requirement of concavity. Optimum source precoding matrix (or equivalently, the covariance matrix) is determined. A special case of precoding, i.e., beamforming, occurs when the covariance matrix is of rank one, which has appealing reduced complexity but may not always be capacity-optimum. Here we derive the necessary and sufficient condition for it to achieve capacity. All our results subsume as a special case those for traditional MISO precoding with channel mean feedback.

\section{System model and problem statement\label{sec_system_model}}
We focus on a half-duplex AF link with \(M\) antennas at the source and with a single antenna at both the relay and the destination,\footnote{The analysis in this paper can be extended to the case when the relay has multiple antennas. Details are not discussed due to space constraint.} resulting in a MISO  source-relay link (backward channel) and a single-input single-output relay-destination link (forward channel). No direct source-destination link exists. The backward channel \(\mathbf{h}_{B}\) is modeled as~\cite{Visotsky_madhow}:
\begin{align}
\mathbf{h}_{B} = \boldsymbol{\mu} + \sqrt{\alpha}\mathbf{h}_{w}, \label{backward_channel_model}
\end{align}
where \(\mathbf{h}_{w}\) represents the scattering and is distributed as \(\mathcal{N}_{c}(0, \mathbf{I}_{M})\) (circularly symmetric complex Gaussian), and \(\alpha\) is a nonnegative scaling constant (\(\alpha\geq0\)). Only the knowledge of  the source-relay channel mean \(\boldsymbol{\mu}\) and \(\alpha\) is provided to the source, which we refer to as \textit{channel mean feedback} (at the source). In the first time slot, the received signal in the source-relay link is given by
\begin{align}
\nonumber r_{1} = \mathbf{h}_{B}^{\dagger}\mathbf{F}\mathbf{x} + n_{1},
\end{align}
where \(\mathbf{x}\) is the source signal with \(\mathbb{E}\{\mathbf{x}\mathbf{x}^{\dagger}\}=\frac{\gamma}{M}\mathbf{I}_{M}\) and \(n_{1}\) is the noise at the relay distributed as \(\mathcal{N}_{c}(0,1)\). The precoding (shaping) matrix \(\mathbf{F}\) is related to the transmit covariance matrix \(\mathbf{Q}\) through \(\mathbf{Q}\triangleq\frac{1}{M}\mathbf{F}\mathbf{F}^{\dagger}\).
A transmit power constraint is imposed on the source:
\(\mathbb{E}\{\text{tr}(\mathbf{F}\mathbf{x}(\mathbf{F}\mathbf{x})^{\dagger})\} = \gamma\text{tr}(\mathbf{Q})\leq \gamma\),
which yields the constraint on the transmit covariance matrix:
\(\text{tr}(\mathbf{Q})\leq 1\).  Denote the amplifying factor of the relay as \(\eta\). Due to noncoherent relaying, a long-term power constraint, \(G\), is imposed on the relay, i.e.,
\begin{align}
G= \mathbb{E}\{|\eta r_{1}|^{2}\} = \eta^{2}[\gamma\boldsymbol{\mu}^{\dagger}\mathbf{Q}\boldsymbol{\mu} + \gamma\alpha\text{tr}(\mathbf{Q})+1]. \label{relay_pwr_constraint}
\end{align}
Thus, \(\eta = \sqrt{G/\left\{1 + \gamma[\boldsymbol{\mu}^{\dagger}\mathbf{Q}\boldsymbol{\mu} +\alpha\text{tr}(\mathbf{Q})]\right\}}\). In (\ref{relay_pwr_constraint}), we have used:
\(\mathbb{E}\{\mathbf{h}_{w}^{\dagger}\mathbf{Q}\mathbf{h}_{w}\}=\text{tr}(\mathbf{Q})\),
since \(\mathbf{h}_{w}\) is \(\mathcal{N}_{c}(0, \mathbf{I}_{M})\).
In the second time slot, the received signal in the relay-destination link is given by
\begin{align}
\nonumber r_{2} = \eta\;h_{F}\;r_{1} + n_{2},
\end{align}
where \(h_{F}\) is the forward channel coefficient, and \(n_{2}\) is the noise at the destination with distribution \(\mathcal{N}_{c}(0,1)\).

Perfect knowledge of \(\mathbf{h}_{B}\) and \(h_{F}\) is assumed at the destination. No knowledge about \(h_{F}\) is available at the source or relay. The ergodic capacity of the above relay channel is given below~\cite{Prathapa_ICC}
\begin{align}
\max_{\substack{\mathbf{Q}\succeq 0\\\text{tr}(\mathbf{Q})\leq 1}} \frac{1}{2}\mathbb{E}_{\mathbf{h}_{B}, \; h_{F}}\left\{\log\left[1+\frac{\eta^{2}\gamma|h_{F}|^{2}\mathbf{h}_{B}^{\dag}\mathbf{Q}\mathbf{h}_{B}}
{\eta^{2}|h_{F}|^{2}+1}\right]\right\},
\label{capacity_expression}
\end{align}
where \(\frac{1}{2}\) is due to the half-duplex assumption. The \(\log\) function here denotes natural logarithm, and thus the unit is nats per channel use. After substituting \(\eta\) into (\ref{capacity_expression}) and applying a technique used in~\cite[Appendix A]{ding_Blostein_PIMRC09}, the ergodic capacity \textit{under mean feedback} can be further shown equivalent to:
\begin{align}
&\max_{\mathbf{Q}\succeq 0,\;\text{tr}(\mathbf{Q})=1} \text{C}(\mathbf{Q}) \label{cmf_problem_formulation}
\end{align}
where
\begin{align}
\text{C}(\mathbf{Q})=\frac{1}{2}\mathbb{E}_{\mathbf{h}_{B},\; h_{F}}\left\{\log\left[1+\frac{G|h_{F}|^{2}\mathbf{h}_{B}^{\dagger}\mathbf{Q}\mathbf{h}_{B}}
{\boldsymbol{\mu}^{\dagger}\mathbf{Q}\boldsymbol{\mu}+\alpha+
\frac{G|h_{F}|^{2}+1}{\gamma}}\right]\right\}. \label{C_Q}
\end{align}

Our goal is to find the optimum \(\mathbf{Q}\) for the problem defined by (\ref{cmf_problem_formulation})-(\ref{C_Q}).
The main challenge here is that  in (\ref{C_Q}), the \(\log\) function inside the expectation operator is nonconvex in \(\mathbf{Q}\). Previous methods for MISO precoding with channel mean feedback~\cite{Visotsky_madhow}-\cite{ALMoustakas}, which generally utilize the concavity of the objective function in the transmit covariance matrix, are not applicable here. Below we employ a new method based on the Laplace transform order~\cite{Shaked_Shanth}. Some mathematical preliminaries pertaining to the Laplace transform order are given in Appendix I.

\section{Optimum Source Covariance Matrix}~\label{solution_proof_cmf}
\textit{Theorem 1} \;\; The optimum \(\mathbf{Q}\) for the problem (\ref{cmf_problem_formulation}) is given by \(\mathbf{Q}_{opt} = \mathbf{V} \boldsymbol{\Phi}\mathbf{V}^{\dagger}\),
where
\begin{align}
\mathbf{V} = &\; \left[\boldsymbol{\mu}/\|\boldsymbol{\mu}\| \hspace{2mm} \mathbf{v}_{2} \; \ldots \; \mathbf{v}_{ M}\right], \label{cmf_opt_eigenvectors}
\\
\boldsymbol{\Phi} =&\; \text{diag}\left\{\phi,  \frac{1-\phi}{M-1}, \ldots,  \frac{1-\phi}{M-1}\right\}, \label{cmf_opt_pwr}
\end{align}
and \(\mathbf{v}_{2} \ldots \mathbf{v}_{ M}\) are arbitrary orthonormal vectors orthogonal to \(\boldsymbol{\mu}\) in \(\mathbb{C}^{M}\).

\textit{Proof}: Due to space constraint, we only provide a detailed outline here. A complete proof can be found in~\cite{mding_kzhang}.

To ease the presentation, we first sketch the basic idea. A globally optimum \(\mathbf{Q}\) exists for (\ref{cmf_problem_formulation}) due to the Weierstrass' Theorem~\cite{Chong}. We divide the feasible set into two subsets, represented by \(\mathbf{Q}_{1}\) and \(\mathbf{Q}_{2}\). Specifically, denote as \(\mathbf{Q}_{1}\) a feasible but otherwise arbitrary covariance matrix, which has \textit{none} of its eigenvectors aligned with \(\boldsymbol{\mu}\). Let \(\mathbf{Q}_{2}\) have the same eigenvectors as \(\mathbf{Q}_{opt}\) [see (\ref{cmf_opt_eigenvectors})]. Here \(\mathbf{Q}_{2}\) is customized for the proof. Our goal is to show that given any \(\mathbf{Q}_{1}\), by proper power allocation in \(\mathbf{Q}_{2}\), we can always have \(\text{C}(\mathbf{Q}_{2})\geq \text{C}(\mathbf{Q}_{1})\). We will show that to maximize \(\text{C}(\mathbf{Q}_{2})\), equal power must be allocated to \(\mathbf{v}_{2} \ldots \mathbf{v}_{ M}\) as in (\ref{cmf_opt_pwr}). We will also show that \(\mathbf{Q}_{2}\) with its optimum power allocation achieves at least the same ergodic capacity as \(\mathbf{Q}_{1}\) does. The optimality of \(\mathbf{Q}_{opt}\) can then be established. Below the proof starts.

To facilitate subsequent comparison, our first step is to exploit the eigen-structures of covariance matrices. Let \(\mathbf{Q}_{1}\) be eigen-decomposed as \(\mathbf{Q}_{1} = \mathbf{U}\boldsymbol{\Lambda}\mathbf{U}^{\dag}\) with
\begin{align}
\mathbf{U} = [\mathbf{u}_{1} \ldots  \mathbf{u}_{M}], \boldsymbol{\Lambda} = \text{diag}\{\lambda_{1},\ldots,\lambda_{M}\}. \label{sub_opt_form}
\end{align}
Here \(\{\mathbf{u}_{1}\ldots \mathbf{u}_{M}\}\) is an arbitrary orthonormal basis in \(\mathbb{C}^{M}\), among which none is aligned with \(\boldsymbol{\mu}\), and \(\sum_{i=1}^{M}\lambda_{i}=1\).
Let \(\mathbf{Q}_{2}\) have the same eigenvectors as given in (\ref{cmf_opt_eigenvectors}), i.e.,
\begin{align}
\mathbf{Q}_{2} = \mathbf{V}\hat{\boldsymbol{\Phi}}\mathbf{V}^{\dag}, \; \hat{\boldsymbol{\Phi}}=\text{diag}\{\hat{\phi}_{1},\hat{\phi}_{2},\ldots,\hat{\phi}_{M}\}, \; \sum_{i=1}^{M}\hat{\phi}_{i}=1. \label{Phi_hat_tmp}
\end{align}
Also define
\(\boldsymbol{\beta} \triangleq (\beta_{1}\ldots\beta_{M})^{T} \triangleq \mathbf{U}^{\dagger}\boldsymbol{\mu}\).
Clearly, \(\|\boldsymbol{\mu}\|=\|\boldsymbol{\beta}\|\). From (\ref{sub_opt_form}) and (\ref{backward_channel_model}), we obtain
 \(\boldsymbol{\mu}^{\dag}\mathbf{Q}_{1}\boldsymbol{\mu} =\sum_{i=1}^{M}\lambda_{i}|\beta_{i}|^{2}\),
and
\begin{align}
\nonumber\mathbf{h}_{B}^{\dagger}\mathbf{Q}_{1}\mathbf{h}_{B} 
= &\;[\boldsymbol{\beta} + \sqrt{\alpha}\breve{\mathbf{h}}_{w}]^{\dagger}\boldsymbol{\Lambda}[\boldsymbol{\beta} + \sqrt{\alpha}\breve{\mathbf{h}}_{w}] =\;\alpha W_{Q1},
\end{align}
where \(\breve{\mathbf{h}}_{w} \triangleq (\breve{h}_{w1} \ldots \breve{h}_{wM})^{T} \triangleq \mathbf{U}^{\dagger}\mathbf{h}_{w}\) has the same distribution as  \(\mathbf{h}_{w}\), \(|\breve{h}_{wi}+\frac{\beta_{i}}{\sqrt{\alpha}}|^{2}\) is a noncentral chi-square random variable of two degrees of freedom with the noncentrality parameter given by \(|\beta_{i}|^{2}/\alpha\), for all \(i\)~\cite[p. 43]{Proakis}, and \(W_{Q1}\triangleq\sum_{i=1}^{M}\lambda_{i}|\breve{h}_{wi}+\frac{\beta_{i}}{\sqrt{\alpha}}|^{2}\).
Similarly, \(\boldsymbol{\mu}^{\dagger}\mathbf{Q}_{2}\boldsymbol{\mu}=\hat{\phi}_{1}\|\boldsymbol{\mu}\|^{2}\), and
\begin{align}
\nonumber &\;\mathbf{h}_{B}^{\dagger}\mathbf{Q}_{2}\mathbf{h}_{B} =\;[\boldsymbol{\mu} + \sqrt{\alpha}\mathbf{h}_{w}]^{\dagger}\mathbf{V}\hat{\boldsymbol{\Phi}}\mathbf{V}^{\dagger}[\boldsymbol{\mu} + \sqrt{\alpha}\mathbf{h}_{w}]
\\\nonumber = &\;[(\|\boldsymbol{\mu}\| \; 0 \ldots 0) + \sqrt{\alpha}\;\hat{\mathbf{h}}_{w}^{\dagger}]\;
\hat{\boldsymbol{\Phi}}\;[(\|\boldsymbol{\mu}\| \; 0 \ldots 0)^{T} + \sqrt{\alpha}\;\hat{\mathbf{h}}_{w}]
\\\nonumber = &\;\alpha W_{Q2},
\end{align}
where \(W_{Q2} \triangleq \hat{\phi}_{1}|\hat{h}_{w1}+\frac{\|\boldsymbol{\mu}\|}{\sqrt{\alpha}}|^{2} + \sum_{i=2}^{M}\hat{\phi}_{i}|\hat{h}_{wi}|^{2}\). Here
\(\hat{\mathbf{h}}_{w} \triangleq (\hat{h}_{w1} \ldots \hat{h}_{wM})^{T} \triangleq \mathbf{V}^{\dagger}\mathbf{h}_{w}\) has the same distribution as  \(\mathbf{h}_{w}\), and thus  \(|\hat{h}_{wi}|^{2}\) is distributed as central chi-square of two degrees of freedom (or simply, exponential), \(i=2, \ldots, M\).

Now, within the subset represented by \(\mathbf{Q}_{2}\), it can be shown that, given any \(0\leq\hat{\phi}_{1}\leq 1\), among all \(\hat{\boldsymbol{\Phi}}\) matrices [see (\ref{Phi_hat_tmp})],
\begin{align}
\hat{\boldsymbol{\Phi}}^{*}=\text{diag}\{\hat{\phi}_{1}, \frac{1-\hat{\phi}_{1}}{M-1},\ldots, \frac{1-\hat{\phi}_{1}}{M-1}\}
\label{opt_hat_Phi}
\end{align}
maximizes \(\text{C}(\mathbf{Q}_{2})\); i.e., to maximize \(\text{C}(\mathbf{Q}_{2})\), equal power  \(\frac{1-\hat{\phi}_{1}}{M-1}\) must be allocated to \(\mathbf{v}_{2}\ldots \mathbf{v}_{M}\)~\cite{mding_kzhang}. Thus, we denote
\begin{align}
\hat{W}_{Q2} \triangleq \hat{\phi}_{1}\left|\hat{h}_{w1}+\frac{\|\boldsymbol{\mu}\|}{\sqrt{\alpha}}\right|^{2} + \frac{1-\hat{\phi}_{1}}{M-1}\sum_{i=2}^{M}|\hat{h}_{wi}|^{2}. \label{S_Q2_tilde}
\end{align}

Up to now, we have
\begin{align}
 \text{C}(\mathbf{Q}_{1}) = &\;\frac{1}{2}\mathbb{E}_{\mathbf{h}_{B}, h_{F}} \{\log[1+k_{1}W_{Q1}]\}, \label{C_Q1}
\\ \text{C}(\mathbf{Q}_{2}) = &\;\frac{1}{2}\mathbb{E}_{\mathbf{h}_{B}, h_{F}} \{\log[1+k_{2}\hat{W}_{Q2}]\}, \label{C_Q2}
\end{align}
where optimum equal power allocation among \(\mathbf{v}_{2}\ldots \mathbf{v}_{M}\) is used in \(\mathbf{Q}_{2}\),
\begin{align}
k_{1} =& \frac{G|h_{F}|^{2}\alpha}{\sum_{i=1}^{M}\lambda_{i}|\beta_{i}|^{2}+ \alpha + \frac{1+G|h_{F}|^{2}}{\gamma}}, \label{k1}
\\k_{2} =& \frac{G|h_{F}|^{2}\alpha}{\hat{\phi}_{1}\|\boldsymbol{\mu}\|^{2}+\alpha+\frac{1+G|h_{F}|^{2}}{\gamma}}. \label{k2}
\end{align}
Note that (\ref{C_Q1}) and (\ref{C_Q2}) differ not only in \(W_{Q1}\) and \(\hat{W}_{Q2}\), but also in \(k_{1}\) and \(k_{2}\), making further comparison prohibitively difficult. To proceed, we choose the only free (unspecified) parameter in \(\mathbf{Q}_{2}\), i.e., \(\hat{\phi}_{1}\), as follows:
\begin{align}
\hat{\phi}_{1} = \frac{\boldsymbol{\mu}^{\dag}\mathbf{Q}_{1}\boldsymbol{\mu}}{\|\boldsymbol{\mu}\|^{2}} =\frac{\sum_{i=1}^{M}\lambda_{i}|\beta_{i}|^{2}}{\|\boldsymbol{\mu}\|^{2}} = \frac{\sum_{i=1}^{M}\lambda_{i}|\beta_{i}|^{2}}{\|\boldsymbol{\beta}\|^{2}},  \label{beta_mu_constraint}
\end{align}
such that \(k_{1} = k_{2} = k > 0\) in (\ref{k1}) and (\ref{k2}).  Since \(0\leq \min_{i}\lambda_{i}\leq\frac{\boldsymbol{\mu}^{\dag}\mathbf{Q}_{1}
\boldsymbol{\mu}}{\|\boldsymbol{\mu}\|^{2}}\leq\max_{i}\lambda_{i}\leq 1\), (\ref{beta_mu_constraint}) is always valid. The choice of \(\hat{\phi}_{1}\) in (\ref{beta_mu_constraint}) is crucial and will be shown to enable the final comparison. Naturally, our next step is to show that
\begin{align}
&\; \mathbb{E}_{W_{Q1}}\{\log(1+ kW_{Q1})\} \leq \mathbb{E}_{\hat{W}_{Q2}}\{\log(1+ k\hat{W}_{Q2})\}, \label{equi_prob2}
\end{align}
when \(h_{F}\) is given and  \(\hat{\phi}_{1}\) is chosen as per (\ref{beta_mu_constraint}). 

A straightforward method to show the above is to calculate the expectations on both sides of the inequality. The difficulty here is that the calculation involves the probability density function (p.d.f.) of a convex combination of \(M\) non-central chi-square random variables, which is too complicated to serve our purpose~\cite{chi_square_convex_combinations}. On the other hand, the Laplace transform of this p.d.f. does possess a more elegant structure~\cite{Proakis}. If we can avoid the p.d.f. and use its Laplace transform instead, we will be able to overcome the difficulty. It turns out that \textit{Lemma 1} in Appendix I is the precise tool we need here.

Based on \textit{Lemma 1}, to show (\ref{equi_prob2}), it suffices to show that  \[W_{Q1}\leq_{\text{LT}}\hat{W}_{Q2}, \; \text{subject to}\; (\ref{beta_mu_constraint}).\] Let \(\mathcal{M}_{W_{Q1}}(s)\) [\(\mathcal{M}_{\hat{W}_{Q2}}(s)\)] be the Laplace transform of the p.d.f. of \(W_{Q1}\) (\(\hat{W}_{Q2}\)). According to \textit{Definition 1} (see Appendix I), it is equivalent to show that \(\mathcal{M}_{\hat{W}_{Q2}}(s)\leq \mathcal{M}_{W_{Q1}}(s), \forall s>0, \text{subject to}\; (\ref{beta_mu_constraint})\),
or,
\begin{align}
\log\;[\mathcal{M}_{\hat{W}_{Q2}}(s)/\mathcal{M}_{W_{Q1}}(s)]\leq 0, \;\forall s>0, \;\text{under}\; (\ref{beta_mu_constraint}). \label{temp_s}
\end{align}
It can be shown that~\cite[p. 43]{Proakis}
\begin{align}
\nonumber &\; \log\left[\mathcal{M}_{\hat{W}_{Q2}}(s)/\mathcal{M}_{W_{Q1}}(s)\right] = \mathcal{J}(s) - \frac{s}{\alpha}\mathcal{R}(s), \\ &\;\mathcal{J}(s) = \log\left[\frac{(1+\lambda_{1}s)\ldots (1+\lambda_{M}s)}{(1+\hat{\phi}_{1}s)\left(1+\frac{1-\hat{\phi}_{1}}{M-1}s\right)^{M-1}}\right] \label{J_s}
\\&\; \mathcal{R}(s) = \frac{\hat{\phi}_{1}\|\boldsymbol{\mu}\|^{2}}{1+\hat{\phi}_{1}s}-
\sum_{i=1}^{M}\frac{\lambda_{i}|\beta_{i}|^{2}}{1+\lambda_{i}s}. \label{R_s}
\end{align}
To show that \(\mathcal{J}(s)\leq 0\) for all \(s>0\), note that \(\sum_{i=1}^{M}\log(1+t_{i}s)\) is a Schur-concave function in \(\mathbf{t} = (t_{1}\ldots t_{M})^{T}\), for all \(s>0\), and subject to (\ref{beta_mu_constraint}),
\[\left(\hat{\phi}_{1} \; \frac{1-\hat{\phi}_{1}}{M-1}\ldots \frac{1-\hat{\phi}_{1}}{M-1}\right)\prec \left(\lambda_{1} \ldots \lambda_{M}\right), \]
i.e., the left-hand side is majorized by the right-hand side~\cite{mding_kzhang}\cite{Marshall_Olkin}. We can also show that \(\mathcal{R}(s)\geq 0\), \(\forall s>0\), by repeatedly using (\ref{beta_mu_constraint}). Thus,  (\ref{temp_s}) holds, which implies that (\ref{equi_prob2}) holds. Since the construction of \(\mathbf{Q}_{2}\) involves only (\ref{cmf_opt_eigenvectors}), (\ref{opt_hat_Phi}) and (\ref{beta_mu_constraint}), none of which depends on \(h_{F}\), and (\ref{equi_prob2}) holds for any \(h_{F}\), we obtain
\begin{align}
\nonumber \text{C}(\mathbf{Q}_{1}) = &\;\frac{1}{2}\mathbb{E}_{h_{F}}\left\{\mathbb{E}_{W_{Q1}}\{\log(1+ kW_{Q1})\} \right\}
\\ \leq&\;\frac{1}{2}\mathbb{E}_{h_{F}}\left\{\mathbb{E}_{\hat{W}_{Q2}}\{\log(1+ k\hat{W}_{Q2})\}\right\}= \text{C}(\mathbf{Q}_{2}). \label{final_1}
\end{align}
Based on the arbitrariness of \(\mathbf{Q}_{1}\) and \((\mathbf{v}_{2}\ldots\mathbf{v}_{M})\), we conclude that the optimum solution, as it exists, must have the same eigen-structure as \(\mathbf{Q}_{2}\). Also, as seen in the proof, equal power allocation among \(\mathbf{v}_{2}\ldots \mathbf{v}_{M}\) is necessary for optimality. At this moment, the only parameter in \(\mathbf{Q}_{2}\) available for further optimization is \(\hat{\phi}_{1}\). Though \(\hat{\phi}_{1}\) chosen as in (\ref{beta_mu_constraint}) is sufficient to guarantee (\ref{final_1}) for a specific \(\mathbf{Q}_{1}\), it can be potentially further optimized to obtain \(\phi\) as in \(\mathbf{Q}_{opt}\). Therefore, for (\ref{cmf_problem_formulation}), \(\mathbf{Q}_{opt}\) [see~(\ref{cmf_opt_eigenvectors})-(\ref{cmf_opt_pwr})] is the optimum with \(\phi\) numerically optimized according to the fading statistics of \(\mathbf{h}_{B}\) and \(h_{F}\). \(\Box\)

\vspace{1mm}

\textit{Remark 1} \;\; The fading distribution of the relay-destination channel \(h_{F}\) has no effect on the optimum transmit directions (eigenvectors of the covariance matrix) at the source. However, it does affect the optimum value of \(\phi\). In fact, \(\phi\) is determined by solving: \(\max_{0\leq\phi\leq1}\frac{1}{2}\mathbb{E}\{\log[1+G|h_{F}|^{2}\alpha\tilde{c}(\phi)]\}\), where
\[\tilde{c}(\phi) =\frac{\phi|\hat{h}_{w1}
+\frac{\|\boldsymbol{\mu}\|}{\alpha}|^{2}+\frac{1-\phi}{M-1}\sum_{i=2}^{M}|\hat{h}_{wi}|^{2}}
{\phi\|\boldsymbol{\mu}\|^{2}+\alpha+\frac{1+G|h_{F}|^{2}}{\gamma}}.\]
This problem can be readily solved using one-dimensional search methods~\cite{Chong}. It is also interesting to see that the capacity depends on \(\boldsymbol{\mu}\) only through its Euclidean length \(\|\boldsymbol{\mu}\|\).

\textit{Remark 2} \;\; When the relay power \(G\rightarrow\infty\), (\ref{cmf_problem_formulation}) becomes:
\(\max_{\mathbf{Q}\succeq 0, \; \text{tr}(\mathbf{Q})=1} \frac{1}{2}\mathbb{E}_{\mathbf{h}_{B}}\{\log[1+\gamma\mathbf{h}_{B}^{\dag}\mathbf{Q}\mathbf{h}_{B}]\}\), which is the same mathematical problem as in~\cite[Theorem 3.1]{Visotsky_madhow}. Thus, our result subsumes as a special case the optimum (traditional) MISO precoding with channel mean feedback, and it is not surprising to see the result in \textit{Theorem 1} and that in~\cite[Theorem 3.1]{Visotsky_madhow} share the same structure.\footnote{Similar observations have also been reported in~\cite{Prathapa_ICC} with channel covariance feedback.} In particular, our proof here can also serve to prove~\cite[Theorem 3.1]{Visotsky_madhow}.

\section{Optimality of Beamforming along \(\boldsymbol{\mu}\)}
Beamforming along the source-relay channel mean \(\boldsymbol{\mu}\) is optimum if and only if all the source transmit power is allocated to \(\boldsymbol{\mu}/\|\boldsymbol{\mu}\|\), and thus \(\phi=1\) and \(\mathbf{Q}_{opt}\) is rank-one.

\textit{Theorem 2} \;\; Assume that \(h_{F}\) is distributed as \(\mathcal{N}_{c}(0,1)\).\footnote{Note that the result in \textit{Theorem 1} holds with any fading distribution of \(h_{F}\). However, the condition for beamforming to be optimum does depend on the distribution of \(h_{F}\).} Given \(\gamma\), \(G\), \(\boldsymbol{\mu}\) and \(\alpha\), beamforming in the direction of \(\boldsymbol{\mu}\) can achieve capacity if and only if
\begin{align}
\nonumber \mathbb{E}\{Z\}+&\frac{1}{G}\mathbb{E}\{Z\exp(Z)\Gamma(0,Z)\}
\\&\hspace{15mm}\leq\mathbb{E}\{Z^{2}\exp(Z)\Gamma(0,Z)\}
+D_{2},\label{BF_condition_ineq}
\end{align}
where the expectation is taken with respect to the random variable \(Z\) with the following probability density function
\begin{align}
\nonumber p_{Z}(z) =
&\frac{D_{1}}{\alpha\gamma z^{2}}\exp\left\{-\left[\frac{\|\boldsymbol{\mu}\|^{2}}{\alpha}+\frac{1}{\alpha\gamma}
\left(\frac{D_{1}}{z}-1\right)\right]\right\}
\\\nonumber\times I_{0}&\left((2\|\boldsymbol{\mu}\|\sqrt{(D_{1}/z)-1})/(\alpha\sqrt{\gamma})\right); \hspace{3mm} 0<z\leq D_{1}.
\end{align}
In the above, we have defined \(D_{1}\triangleq (\alpha\gamma +1 +\gamma\|\boldsymbol{\mu}\|^{2})/G\) and \(D_{2}\triangleq \frac{D_{1}}{\alpha\gamma+1}\left[1-\frac{\gamma\|\boldsymbol{\mu}\|^{2}}{G}\exp(D_{1})\Gamma\left(0,D_{1}\right)\right]\), \(\Gamma(a, x)\) is the complementary incomplete Gamma function~\cite[Eqs. (6.5.3), (6.5.15)]{integration_book}, and \(I_{0}(x)\) is the zeroth-order modified Bessel function of the first kind~\cite[Eqs. (9.6.10), (9.6.16)]{integration_book}.

\textit{Proof}: Due to space limitation, the proof is omitted.
Numerical methods are required to evaluate \(\mathbb{E}\{Z\}\), \(\mathbb{E}\{Z\exp\{Z\}\Gamma(0,Z)\}\) and \(\mathbb{E}\{Z^{2}\exp\{Z\}\Gamma(0,Z)\}\). \(\Box\)

\textit{Remark 3} \;\; When \(G\rightarrow\infty\), (\ref{BF_condition_ineq}) coincides with~\cite[Theorem 4, \(n_{R}=1\)]{Jafar_Goldsmith}.

\section{Numerical Examples}~\label{sec_numerical_examples}
\begin{figure}[!t]
\begin{center}
\includegraphics[height=5in,width=3.5in, keepaspectratio]
{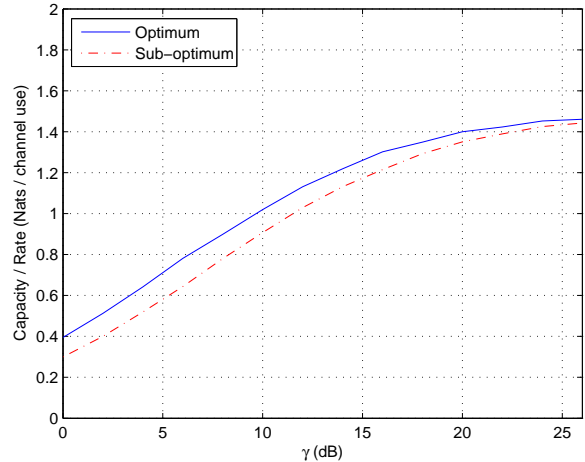} \caption{Comparison of optimum and sub-optimum solutions for $M=2$, $\alpha = 0.1$,  $\boldsymbol{\mu}=(0.3518 + j 0.2496 \; -0.4039 - j 1.0437)^{T}$ ($\|\boldsymbol{\mu}\|^{2}/\alpha = 14.3851$),  $G = 15$ dB.} \label{fig_1}
\end{center}
\end{figure}

\begin{figure}[!t]
\begin{center}
\includegraphics[height=5in,width=3.5in, keepaspectratio]
{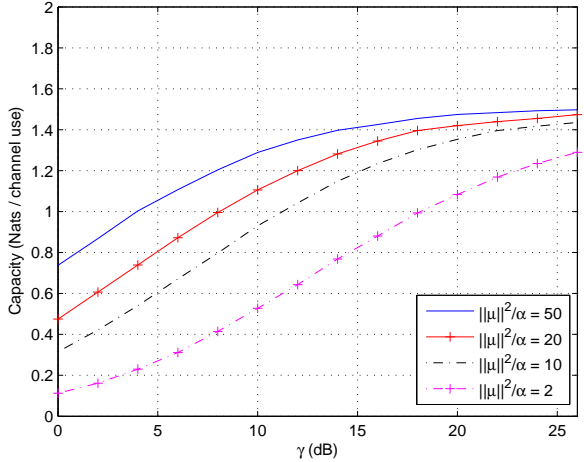} \caption{Illustration of the impact of $\|\boldsymbol{\mu}\|$ on capacity for $M=2$, $\alpha = 0.1$, $G = 15$ dB.} \label{fig_2}
\end{center}
\end{figure}

We now provide simulation results to corroborate the analytical results.  We choose the number of antennas at the source \(M\) to be 2. Fig.~\ref{fig_1} shows capacity versus \(\gamma\). Since the noise power is normalized to one, \(\gamma\) denotes the transmit signal-to-noise ratio (SNR). Here the ``optimum'' refers to \(\mathbf{Q}_{opt}\) which achieves capacity, and the ``sub-optimum'' refers to a sub-optimum \(\mathbf{Q}_{1}\) (see the proof of \textit{Theorem 1}) with numerically optimized power allocation. The optimality of \(\mathbf{Q}_{opt}\) is clearly shown in Fig.~\ref{fig_1}. At high \(\gamma\), the difference between the rates using the optimum and the sub-optimum diminishes. From Fig.~\ref{fig_2}, capacity increases with \(\|\boldsymbol{\mu}\|\) with other parameters fixed. Similar observations can be made with different sets of parameters, which are not presented here due to space constraint.

\begin{figure}[!t]
\begin{center}
\includegraphics[height=5in,width=3.5in, keepaspectratio]
{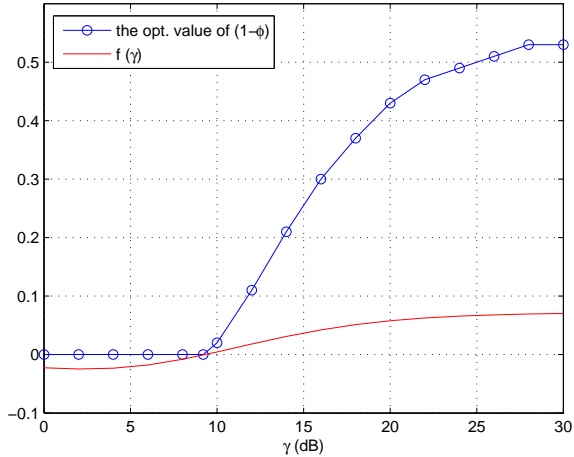} \caption{Illustration of the validity of (\ref{BF_condition_ineq}) for $M=2$, $\alpha = 0.5$,  $\boldsymbol{\mu}=(-0.2163 + j 0.0627 \; -0.8328 + j 0.1438)^{T}$,  $G= 10$ dB. The function $f(\gamma)$ is given in Section~\ref{sec_numerical_examples}.} \label{fig_3}
\end{center}
\end{figure}

Fig.~\ref{fig_3} gives simulation results to corroborate (\ref{BF_condition_ineq}). Consider the case with two antennas at the source (\(M=2\)). Given the parameters $\alpha$, $\boldsymbol{\mu}$, and \(G\), the optimum \(\phi\) can be determined for a specific \(\gamma\), and let \(f(\gamma)= \mathbb{E}\{Z\}+\frac{1}{G}\mathbb{E}\left\{Z\exp(Z)\Gamma(0,Z)\right\}-\left(\mathbb{E}\left\{Z^{2}\exp(Z)\Gamma(0,Z)\right\}
+D_{2}\right)\). According to (\ref{BF_condition_ineq}), if beamforming is optimum, then \(\phi=1\) [or, \(1-\phi\) = 0], and \(f(\gamma)\leq 0\).
This consistency is clearly reflected in Fig.~\ref{fig_3}.

\section{Concluding remarks}~\label{sec_conclusion}
The optimum source covariance matrix of a noncoherent half-duplex AF MISO relay channel has been determined with channel mean feedback at the source. We have used a new method based on the Laplace transform order of two nonnegative random variables. Our results subsume as an asymptotic case the optimum precoding for a traditional MISO link. The superiority of the optimum transmit strategy over sub-optimum ones has been shown by simulations and is seen more pronounced at low to medium transmit SNR. Necessary and sufficient condition for optimality of beamforming has also been derived. It is expected that the powerful Laplace transform ordering approach used in this paper will find many applications pertaining to stochastic optimization problems in wireless communications and signal processing~\cite{mding_qtz_icassp11}.

\section*{Appendix I}
We introduce the key mathematical elements of this paper.

\textit{Definition 1}~\cite[p. 95]{Shaked_Shanth} \;\; Let \(T_{1}\) and \(T_{2}\) be two \textit{nonnegative} random variables such that
\(\mathbb{E}\{e^{-sT_{1}}\}\geq\mathbb{E}\{e^{-sT_{2}}\}, \forall s>0 \).
Then \(T_{1}\) is said to be \textit{smaller} than \(T_{2}\) \textit{in the Laplace transform order}, denoted by \(T_{1}\leq_{\text{LT}} T_{2}\).

\textit{Definition 2}~\cite[p. 96]{Shaked_Shanth} \;\; A function \(q\): \(\mathbb{R}_{+}\rightarrow \mathbb{R}\) is said to be completely monotone if all its derivatives \(q^{(n)}\) exist and \((-1)^{n}q^{(n)}(x)\geq 0\), for all \(x>0\) and \(n=0,1, 2,\ldots\) (all nonnegative integer values).

\textit{Lemma 1} \;\; Let \(T_{1}\) and \(T_{2}\) be two \textit{nonnegative} random variables, and let \(d\) be any positive constant (\(d>0\)). If  \(T_{1}\leq_{\text{LT}}T_{2}\), then \(\mathbb{E}\left\{\log (1 + d T_{1})\right\}\leq \mathbb{E}\left\{\log (1 + d T_{2})\right\}. \)

\textit{Proof}: The proof involves Theorem 3.B.4 (a) (p. 97) and Eq. (3.B.2) (p. 96) of~\cite{Shaked_Shanth}, and is based on the fact that \(\log(1+dx), d>0,\) is a positive function in \(x\) when \(x> 0\) with its first-order derivative being completely monotone (see \textit{Definition 2}). Details can be found in~\cite{mding_kzhang}. \(\Box\)

\section*{Acknowledgment}
The authors would like to thank Dr. P. Dharmawansa, Prof. R. K. Mallik, Prof. M. R. McKay, and Prof. K. B. Letaief for helpful discussions.

\balance



\begin{thebibliography}{}
\bibitem{Laneman_1} J. N. Laneman, D. Tse, and G. W. Wornell, ``Cooperative diversity in wireless
networks: efficient protocols and outage bahavior,'' {\em IEEE Trans. Inf. Theory},
vol. 50, no. 12, pp. 3062-3080, Dec. 2004.

\bibitem{Nabar_performance} H. B\"{o}lcskei, R. U. Nabar, O. Oyman, and A. Paulraj, ``Capacity scaling laws
in MIMO relay networks,'' {\em IEEE Trans. Wireless Commun.}, vol. 5, no.
6, pp. 1433-1444, Jun. 2006.

\bibitem{tang_hua_perfect_CSI} X. Tang and Y. Hua, ``Optimal design of non-regenrative MIMO wireless relays,'' {\em IEEE Trans. Wireless Commun.}, vol. 6, no. 4, pp. 1398-1407, Apr. 2007.

\bibitem{medina_Olga} O. Mun\~{o}z-Medina, J. Vidal, and A. Augst\'{i}n,  ``Linear transceiver design in nonregenerative relays with channel state information,'' {\em IEEE Trans. Signal Process.}, vol. 55, no. 6,
pp. 2593-2604, Jun. 2007.

\bibitem{Prathapa_ICC} P. Dharmawansa, M. R. McKay, R. K. Mallik, and K. B. Letaief,  ``Optimality of beamforming for a correlated MISO relay channel,'' in {\em Proc. IEEE ICC 2010}, pp. 1-5, May 2010.

\bibitem{Dharmawansa_McKay_journal} P. Dharmawansa, M. R. McKay, R. K. Mallik, and K. B. Letaief, ``Ergodic capacity and beamforming optimality for MISO relaying with statistical CSI,'' submitted to {\em IEEE Trans. Commun.}.

\bibitem{MIMO_capacity_goldsmith} A. Goldsmith, S. A. Jafar, N. Jindal, and S. Vishwanath, ``Capacity limits of MIMO channels,'' {\em IEEE JSAC},
vol. 21, no. 5, pp. 684-702, Jun. 2003.

\bibitem{Zhou_Giannakis_CMF} S. Zhou and G. B. Giannakis, ``Optimal transmitter eigen-beamforming and space-time block coding based on channel mean feedback,'' {\em IEEE Trans. Signal Process.}, vol. 50, no. 10, pp. 2599-2613, Oct. 2002.

\bibitem{Visotsky_madhow} E. Visotsky and U. Madhow, ``Space-time transmit precoding with imperfect feedback,'' {\em IEEE Trans. Inf. Theory}, vol. 47, no. 6, pp. 2632-2639, Sep. 2001.

\bibitem{Jafar_Goldsmith} S. A. Jafar and A. Goldsmith, ``Transmitter optimization and optimality of beamforming for multiple antenna systems with imperfect feedback,'' {\em IEEE Trans. Wireless Commun.}, vol. 3, no. 4, pp. 1165-1175,  Jul. 2004.

\bibitem{ALMoustakas} A. L. Moustakas and S. H. Simon,  ``Optimizing multiple-input single-output (MISO)
communication systems with general Gaussian
channels: nontrivial covariance and nonzero mean,'' {\em IEEE Trans. Inf. Theory}, vol. 49, no. 10, pp. 2770-2780, Oct. 2003.

\bibitem{Boche_covariance_feedback} E. Jorswieck and H. Boche, ``Channel capacity and capacity-range of beamforming in MIMO wireless systems under correlated fading with covriance feedback,'' {\em IEEE Trans. Wireless Commun.}, vol. 3, no. 5, pp. 1543-1553, Sep. 2004.
    
\bibitem{mding_sdb_T_IT_2010_to_appear} M. Ding and S. D. Blostein, ``Maximum mutual information design for MIMO systems with imperfect channel knowledge,'' {\em IEEE Trans. Inf. Theory}, vol. 56, no. 10, pp. 4793-4801, Oct. 2010.


\bibitem{mding_kzhang} M. Ding and Q. T. Zhang, ``Source optimization in a noncoherent relay channel with channel mean feedback,'' submitted to {\em IEEE Trans. Inf. Theory} in Jun. 2010.

\bibitem{chi_square_convex_combinations} S. J. Press, ``Linear combinations of non-central chi-square variates,'' {\em Ann. Math. Stat.}, vol. 37, no. 2, pp. 480-487, Apr. 1966.

\bibitem{Proakis} J. G. Proakis, {\em Digital Communications}, McGraw-Hill, 2000.

\bibitem{Shaked_Shanth} M. Shaked, J. G. Shanthikumar, {\em Stochatic Orders and Their Applications}, Academic Press, 1994.

\bibitem{Marshall_Olkin} A. W. Marshall, I. Olkin, {\em Inequalities: Theory of Majorization and Its Applications}, Academic Press, 1979.

\bibitem{Chong} E. K. P. Chong, S. H. \.{Z}ak, {\em An Introduction to Optimization}, 3rd Edition, Wiley, 2008.

\bibitem{integration_book} M. Abramowitz,  I. A. Stegun, {\em Handbook of Mathematical Functions}, Dover Publications, 1965.

\bibitem{ding_Blostein_PIMRC09} M. Ding, S. D. Blostein, et al., ``A general framework for MIMO transceiver design with imperfect CSI and transmit correlation,'' in {\em Proc. IEEE PIMRC}, pp. 182-186, Sep. 2009.
    
 \bibitem{mding_qtz_icassp11} M. Ding and Q. T. Zhang, ``Stochastic optimization based on the Laplace transform order with applications to precoder designs,'' to appear in {\em Proc. IEEE ICASSP 2011}, May 2011.
\end{thebibliography}
\end{document}